\documentclass[referee,onecollarge,12pt]{svjour2}       
\usepackage{graphicx}
\usepackage{rotating}
\begin{document}
\title{Nonmonotonic dependence of the absolute
entropy on temperature in supercooled Stillinger-Weber
silicon}
\author{Pankaj A. Apte and Arvind K. Gautam}
\institute{Department of Chemical Engineering,
         Indian Institute of Technology Kanpur,
         Kanpur,
         U.P, India  208016}
\date{Accepted : 12 September 2012 (Journal of Statistical Physics)}
\maketitle
\begin{abstract}
Using a recently developed thermodynamic integration method, 
we compute
the precise values of the excess Gibbs free energy ($G^e$) 
of the high density
liquid (HDL) phase with respect to the crystalline phase at different
temperatures ($T$) in the supercooled region of the 
Stillinger-Weber (SW) silicon [F. H. Stillinger and 
T. A. Weber, Phys. Rev. B. {\bf 32}, 5262 (1985)].  
Based on the slope of $G^e$ with respect to $T$,
we find that the absolute entropy of the HDL phase increases
as its enthalpy changes from the equilibrium value at $T \ge 1065$ K
to the value corresponding to a 
non-equilibrium state at $1060$ K.
We find that the volume distribution
in the equilibrium HDL phases
become progressively broader as the temperature is
reduced to 1060 K,
exhibiting van-der-Waals (VDW) loop in the
pressure-volume curves.  
Our results provides insight into the thermodynamic cause of the
transition from the HDL phase to the
low density 
phases in SW silicon, observed in earlier 
studies
near 1060 K at zero pressure.
\end{abstract}
\section{Introduction}

The liquid-amorphous transition in silicon, 
modeled by the Stillinger-Weber (SW)
potential~\cite{STILLINGER85},
has been intensely 
studied~\cite{BROUGHTON87,LUEDTKE88,LUEDTKE89,ANGELL96,SASTRY03,BEAUCAGE05,VASISHT11,HUJO11,LIMMER11}
with an aim of understanding the phase behavior of real silicon.  In the initial
molecular dynamics (MD) studies 
on SW silicon~\cite{LUEDTKE89,ANGELL96},
it was found that 
the high density liquid (HDL) phase,
at a sufficiently slow cooling rate,
undergoes a sudden transition to a low density amorphous 
phase at around 1060 K.  The nature of the low density 
phase (i.e., whether it is
a solid or a liquid) below the transition
temperature was however not clear.  
In 2003, Sastry and Angell, through precise and careful 
measurements  of the diffusivity in MD simulations, 
showed that 
a low density liquid (LDL) phase exists 
below 1060 K and hence the transition should be
characterized as a liquid--liquid transition~\cite{SASTRY03}.
It was also demonstrated that
in constant pressure--constant enthalpy (NPH) MD simulations starting
from the HDL phase at  $T > 1060$ K, the enthalpy shows a nonmonotonic
dependence on temperature, ultimately leading to the formation of the
LDL phase~\cite{SASTRY03}.
This was attributed to the release of latent
heat during the phase transformation 
from the HDL phase to the LDL phase~\cite{SASTRY03}.
Recently, studies by
Hujo et. al.~\cite{HUJO11} and Limmer and Chandler~\cite{LIMMER11}
do not suggest the presence of a phase coexistence temperature between
the HDL and the LDL phases at zero pressure.

In this work, we study equilibrium HDL phases at and above 1060 K in
isothermal--isobaric (NPT) Monte Carlo (MC) simulations at zero pressure, 
focusing
particularly on the volume (or density) distributions.  
We find that due to shallow free energy barriers, complete equilibration
of the HDL phase cannot be achieved in some MC trajectories,
leading to non-equilibrium states at 1060 K.
We have used a recently developed
thermodynamic integration method~\cite{GROCHOLA04,APTE06-1,APTE06-2,APTE10}
to measure precisely 
the excess Gibbs free energy ($G^e$)
of the HDL phases with respect to the
crystalline phases at a given temperature and pressure.  
These computations yield information about the
entropy changes in the HDL phases as the temperature is decreased to 1060 K.
Our work provides further insight into the transition
from the HDL phase to the low density phases near 1060 K.
In what follows,
we describe the details of our computational method.  

\section{Equilibration of the HDL phase}

To compute the excess free energy (to be described in the next section), 
it is important to correctly
determine the average properties of the HDL phase in the supercooled region.
To this end, we studied the properties of the supercooled 
HDL phase by performing isothermal-isobaric (NPT)
Monte Carlo (MC) simulations at and above $1060$ K and zero pressure.  
All of our computations were
performed with a system of $N=512$ particles in a cubic simulation box under
periodic boundary conditions.  
Our main result is strongly dependent on the properties of the HDL phase
in the temperature range of 1060--1070 K and hence we focus particularly
on this temperature range.  
We find that at $T < 1086$ K, the  
simulations starting in the HDL phase undergo a transition relatively quickly 
to the low density phases, indicating shallow free energy barriers. 
By trial and error, we find the trajectory that stays in the high density
region for the largest number of MC steps  at 1060, 1065 and 1070 K,
by starting from different initial configurations   
(see  Figs.~\ref{fig:trajectory1} and ~\ref{fig:trajectory2}).
At 1060 K, we also found shorter trajectories (seen in the
inset of Figs.~\ref{fig:trajectory1} and ~\ref{fig:trajectory2}) 
corresponding to  non-equilibrium states.
The computed average densities of the HDL phases (see Table~\ref{tab:table1})
at $T > 1070$ K 
agree reasonably well with those found in the MD cooling experiments of 
Beaucage and Mousseau~\cite{BEAUCAGE05}.  

To obtain the average properties of the HDL phases, 
we consider the entire length of
the trajectory before it exhibits a systematic and
continuous decrease in energy and density (as indicated by the arrow positions
in  Figs.~\ref{fig:trajectory1}
and~\ref{fig:trajectory2}), which signals the crossing of the free energy
barrier. 
One noticeable feature is that these trajectories do not seem to
have converged to the mean values (at the arrow positions), unlike the
metastable states which are normally encountered.
This may be due to the fact that
the probability distribution with respect to the volume and energy
is broad and highly asymmetric for the HDL phases.
At 1065 K and 1070 K, we find that shorter 
trajectories (not shown) generated independently  
(with lengths of 10.4 and 35.2 million MC steps, 
respectively)
acquire average energies and densities 
(just before crossing the free energy barrier),
which are nearly equal to the values
reported here for the longer trajectories.  For this
reason, it is important to consider the
entire length of the trajectories upto the arrow positions
to compute the average properties.  If one considers smaller
portions of the trajectories, the resulting average properties
will not reflect the correct sampling of the free energy surface.

At 1060 K, we find that the average properties of the shorter 
trajectories (denoted 
as 1060-S1 and 1060-S2)
and the longer trajectory (see Table~\ref{tab:table1}) 
differ considerably.
Due to its shallowness,
the free energy barrier is crossed even before the equilibration
is achieved, leading to the non-equilibrated shorter trajectories.
In case of the longer
trajectory at 1060 K  (see
Figs.~\ref{fig:trajectory1}
and~\ref{fig:trajectory2}), the cumulative averages show
an overall drift to lower
densities and energies and therefore
it may appear that the high energy (and high density)
states become inaccessible as the trajectory progresses.
However it is clear from the 
{\it instantaneous}  block averages along the
trajectory in Fig.~\ref{fig:trajectory3}, 
that the higher energy (and
higher density) states remain accessible even towards the end of 
the HDL portion of the trajectory.
The same is the case with the trajectories
at 1065 and 1070 K.  In a latter section, our analysis 
based on Gibbs Helmholtz equation indicates
that trajectories  at 1065 and 1070 K, as well as the longer 
trajectory at 1060 K
represent equilibrium phases.

Next we analyze 
the pressure (p)--volume (v) curves (Fig.~\ref{fig:spinodal1}) and
the Helmholtz free energy of the HDL phases as a function of volume
(Fig.~\ref{fig:fv1065}) from the data generated by the above HDL 
trajectories.   Here $v = V/N$ is the volume per
particle (in units of $\sigma^3$)
and $p$ is the average virial
pressure (in units of $\epsilon/\sigma^3$)
obtained from fluctuations that correspond to a bin of 
width $\Delta v = 6.24 \times 10^{-4} \sigma^3$ 
taken at $v$.  (The quantities
reported throughout this work are expressed in units of
SW potential parameters~\cite{STILLINGER85} $\sigma$ and $\epsilon$,
unless otherwise noted explicitly).  We obtain the Helmholtz free 
energy at a given v by using the fact that in isothermal--isobaric
ensemble at zero pressure,
the probability density of finding the system with a specific 
volume v is
$\exp[-\beta F(V)]$.  Hence if
$N_c$ is the number of 
configurations generated in MC simulations with a specific volume
between $v$ and $v+\Delta v$, then
$N_c \propto \exp[-\beta F(V)] \Delta v$ and hence
$\log N_c = -\beta F +$ constant.
This is the quantity we have plotted verses v in 
Fig.~\ref{fig:fv1065} and also in the subsequent figures.
  
At 1070 K, we observe a region with an approximately zero slope in 
the p-v curve as indicated
by the two highlighted points along the curve in
Fig.~{\ref{fig:spinodal1}}, indicating the presence of a two-phase region.   
At 1060 and 1065 K, we observe a region in the p-v curve with a
positive slope.  
At 1060 K and 1065 K, we are able to construct double tangent lines
corresponding to the 
Van-der-Waals (VDW) 
loops as seen in Fig.~\ref{fig:double_tangent}.
The two ends of the
tangent line represents two
states with the same chemical potential at the same temperature
and pressure and hence the 
relation $\Delta U + p \Delta V = T \Delta S$ 
is valid~\cite{CALLEN85}, where $\Delta$ represents the 
difference of the properties of the two states across the tangent line.  
In the present case,
the values of $p$ in the above equation are $-0.65 \times 10^{-3}$
and $-0.27 \times 10^{-2}$ at 1060 K (long trajectory) and 1065 K, 
respectively.  Both the terms $\Delta U$ and $p \Delta V$ have the same sign
indicating a non-zero enthalpy difference between states joined by the
double-tangent lines (see Fig.~\ref{fig:double_tangent}).

At 1065 K, we find that F-v curve 
is not symmetric about the maximum even for small 
deviations away from the maximum.  Instead we find that the
distribution 
can be described fairly accurately by a Taylor series
expansion around the spinodal according to the following equation.
\begin{equation}
   F (T,V,N) = F_s - p_s (V-V_s)
               +\frac{1}{3!}F_s'''(V-V_s)^3,
\label{eq:prob_density3}
\end{equation}
Here $V_s$ corresponds to the volume at the spinodal.  The second order term
is taken as zero, since by definition, the second derivative is zero at the
spinodal.  Taking derivatives by finite difference always involves much noise 
and therefore it is not possible to reliably compute $F_s'''$ directly
by numerical differentiation.  
We used the actual value of $p_s$ (average virial pressure)
as measured in simulation at the
spinodal density (see Fig.~\ref{fig:double_tangent}), 
while we fitted the value of $F_s'''$ by trial and error
so as to best fit the data.  The resultant curve agrees well with the actual
value of $F$ as seen in Fig.~\ref{fig:sp1065}.  We considered 
$p_s = -2.935 \times 10^{-3}$ and $F_s''' = -2.872 \times 10^{-5}$ for the
left spinodal at $v_s = 2.1111$ and 
$p_s = -2.352 \times 10^{-3}$ and $F_s''' = 5.961 \times 10^{-5}$ for the
right spinodal at $v_s = 2.1205$.
This suggests that the probability distribution with respect to $V$,
is controlled
by the spinodals at the two ends of the unstable region.  

Also, we find that the expansion on either side of the unstable region
is accurate only upto the spinodal and deviates rapidly from the actual curve
on the other side of the spinodal as shown in the inset of Fig.~\ref{fig:sp1065}.  
This implies a discontinuity in the equation of state at the two spinodals, which 
is expected because the two ends of the VDW loop usually correspond to the 
two separate phases
each having its own equation of state and 
that each phase extends right upto the spinodal
on either side of the VDW loop.  
It is to be noted that the HDL phase configurations 
consists of atoms connected to form tetrahedra~\cite{BEAUCAGE05}.  
The different tetrahedra 
are connected through common atoms forming a network 
and the number of these tetrahedral structures
can fluctuate causing overall variations 
in the potential energy and the density.  
The above
Taylor series expansion around the spinodals 
may be attributed to the network forming tendency
of the liquid.  This needs to be investigated further.

We observe
that the following fluctuation
relation~\cite{LANDAU-5} is satisfied by the MC trajectories:
\begin{equation}
    \langle (\Delta p)(\Delta V) \rangle = - k_B T,
\label{eq:dpdv}
\end{equation}
when we consider $\Delta p = (p_{in}-\langle p_{in} \rangle)$ and  
$\Delta V =(V - \langle V \rangle) $,
where $p_{in}$ is the instantaneous pressure in the NPT-MC simulations
calculated from the virial relation~\cite{ALLEN87} 
at the
given instantaneous volume $V$ and $T$ is the externally set temperature.  
The symbol $\langle \cdots \rangle$ represents the
average taken over the entire trajectory of the isothermal
isobaric MC simulations.  
This fluctuation relation is derived 
by assuming~\cite{LANDAU-5}
that the instantaneous fluctuations represents a change in 
state from one homogeneous phase to the other.  The relation is satisfied by
the entire trajectory consisting of the HDL, LDL and the defect crystal regions,
as well as by the partial trajectories consisting only of the HDL phase region.
This possibly indicates that transition from the HDL phase to 
the LDL or the d-crystal 
phases along the trajectory 
involves entirely homogeneous states.
We also find that the total average virial 
pressure is zero at all points along the MC trajectories meaning that the
system is in mechanical equilibrium throughout.

\section{Computation of excess Gibbs free energy}

\begin{figure}
  \begin{center}
    \includegraphics[height=12 cm,width=10 cm]{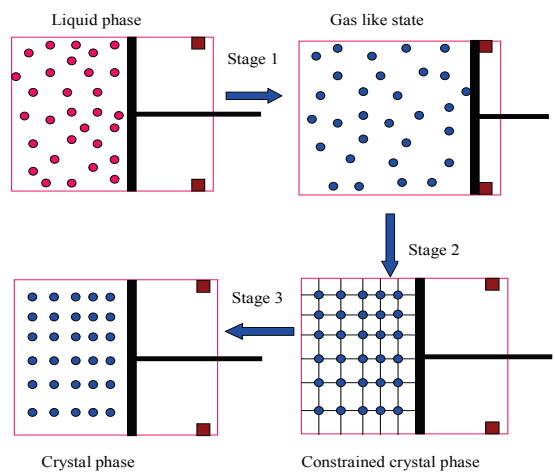}
  \end{center}
   {
  \caption{\label{fig:method} 
   Schematic diagram of the thermodynamic integration path connecting the liquid
   and the solid states  at constant
   temperature and constant external pressure.~\cite{GROCHOLA04,APTE06-1}
   The stops in the piston-cylinder arrangement represent the maximum volume
   constraint as described in the text.
   }}
\end{figure}

We computed the excess Gibbs free energy difference 
$G^{e} =G_{\mbox{\small HDL}} - G_{\mbox{\small crystal}}$ 
between the HDL and the crystal phases, 
by applying the constrained fluid $\lambda$ integration method~\cite{GROCHOLA04} in
the isothermal--isobaric ensemble~\cite{APTE06-1,APTE06-2,APTE05}.  Recently, 
the method was found to predict the melting point of SW silicon accurately~\cite{APTE10}.
This is a thermodynamic integration method in which the liquid and the crystal
phase are connected directly through a 
3-stage reversible path.  
In stage 1 of the reversible path, which starts from  the liquid phase,
the strength of the interaction potential is reduced linearly so that the system approaches
an ideal gas-like behavior.   The expression for the potential energy in this stage is given
by~\cite{GROCHOLA04,APTE06-1},
\begin{equation}
    \phi_1(\lambda_1) = (1- \eta \lambda_1) \phi,
\label{eq:phi1}
\end{equation}
where $\eta$ is a constant that determines the effective strength of the interaction potential
at the end of stage 1 and $\phi$ is the original inter-particle potential 
(SW potential, in the present case).  The parameter $\lambda_1$ defines the states along the
path and varies from 0 to 1. 
As the system becomes less attractive it tends to expand.  However, 
to maintain the reversibility of the path in stages 2 and 3, it is necessary to impose a maximum
volume constraint~\cite{APTE06-1}.
The maximum constrained volume $V_m$ is chosen such that it
is slightly larger than the average volumes of the liquid and the 
solid phase (whichever is larger).  At
the same time, $V_m$ should not affect the free energies 
of either of these phases~\cite{APTE06-1}.  Such a volume
can be straightforwardly chosen based 
on the histogram of volume fluctuations for the two phases.
As in an earlier work on SW silicon~\cite{APTE10}, we chose 
$V_m$ to correspond to a density of $0.4 \sigma^{-3}$, i.e., $V_m = N/0.4$.
At the end of stage 1, we get a compressed gas phase due
to the constraint on the maximum volume~\cite{APTE06-1}.
This process is depicted pictorially in Fig.~\ref{fig:method}.

In stage 2, we force the particles to form a crystalline structure by imposing
an external potential in the form of Gaussian potential wells distributed on the 
ideal crystal
lattice~\cite{GROCHOLA04}.
The strength of the
inter-particle potential energy is held fixed during this stage.  The total
potential energy for stage 2 is given by~\cite{GROCHOLA04,APTE06-1}
\begin{equation}
   \phi_2(\lambda_2) = (1-\eta) \phi + \lambda_2 U_{ext}.
	\label{eq:phi2}
\end{equation}
The Gaussian external potential imposed during this and the subsequent stage is given 
by $U_{ext} = \sum_{i} \sum_{k} a \exp(-b r_{ik}^2)$~\cite{GROCHOLA04}.
Here, the index `i' refers to
the system particle and the index `k' is a Gaussian potential well.  The Gaussian well
does not act on a specific particle, but exerts an influence over all the particles 
in its vicinity.~\cite{GROCHOLA04,APTE06-1} The values of the
Gaussian parameters are taken to be the same as in the earlier work~\cite{APTE10}:
$\eta = 0.9$, $a=-1.892 \epsilon$ and $b=8.0 \sigma$.  These values are so chosen that
the constrained crystalline state obtained at the end of stage 2 has almost the
same energy and density as the desired crystal phase~\cite{APTE05,APTE06-1}.

In stage 3, the Gaussian external potential is reduced linearly to zero, while the strength
of the potential energy is increased linearly to its original value~\cite{GROCHOLA04}.
The potential energy expression for this stage is given by~\cite{GROCHOLA04}
\begin{equation}
 \phi_3(\lambda_3) = [(1-\eta)+\lambda_3 \eta] \phi + (1-\lambda_3) U_{ext}.
\label{eq:phi3}
\end{equation}
At the end of this stage, we get the desired crystalline phase as shown
pictorially in Fig.~\ref{fig:method}.

\begin{figure}
  \begin{center}
    \includegraphics[height=7.1cm,width=8.21cm]{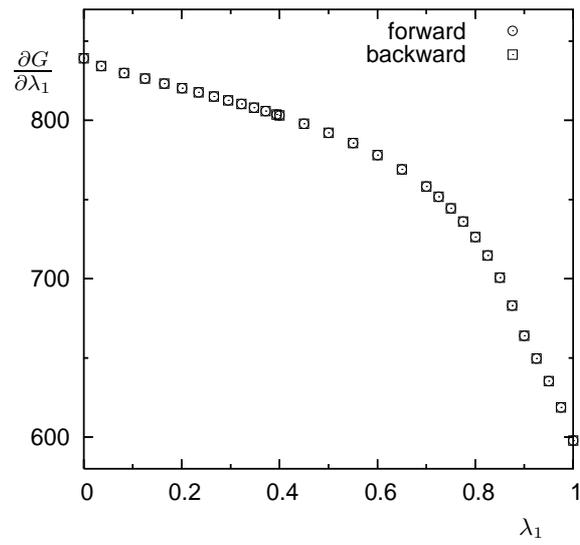}
    \put(-225,165){\large 
                     $\frac{\partial G} {\partial \lambda_1}$ 
                   }
    \put(-38,-7){ {$\lambda_1$}}
  \end{center}
   {
  \caption{\label{fig:s1}
     The integrand (in units of the SW potential parameter $\epsilon$)
     in Eq.~(\ref{eq:dgi}) as a function of $\lambda_1$ for stage 1 at $1065$ K,
     $P = 0$ and $N = 512$.
   }}
\end{figure}

\begin{figure}
  \begin{center}
    \includegraphics[height=7.1cm,width=8.21cm]{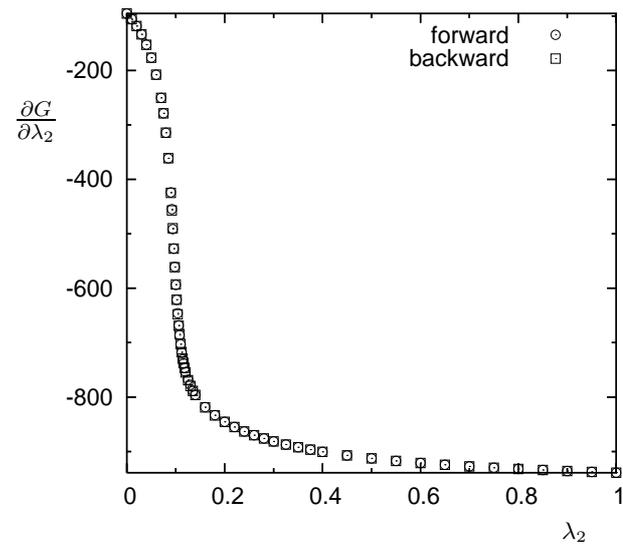}
    \put(-240,148){\large 
                     $\frac{\partial G} {\partial \lambda_2}$ 
                   }
    \put(-38,-7){ {$\lambda_2$}}
  \end{center}
   {
  \caption{\label{fig:s2}
     The same as in Fig.~\ref{fig:s1}, but for stage 2.
   }}
\end{figure}

\begin{figure}
  \begin{center}
    \includegraphics[height=8.5cm,width=9 cm]{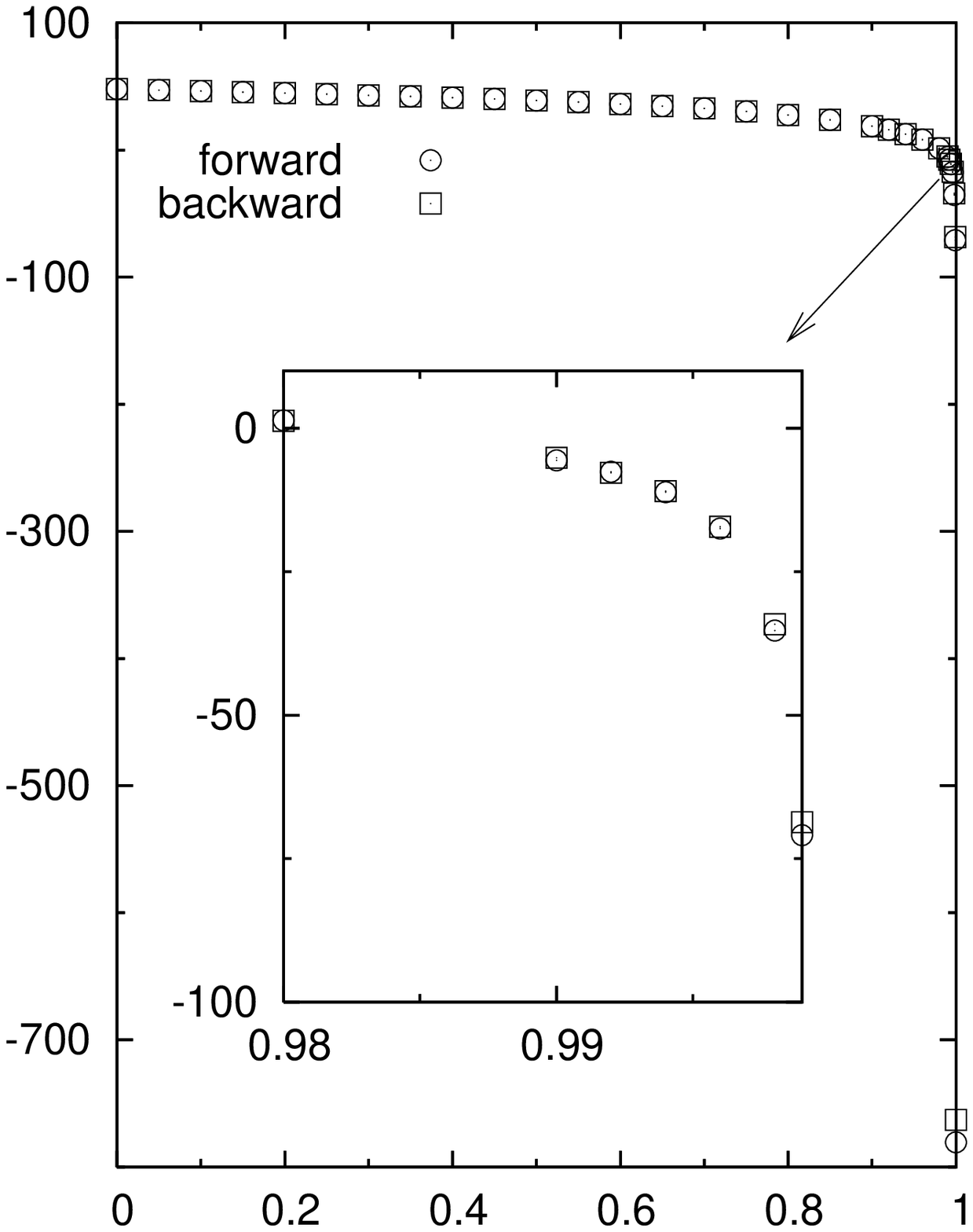}
    \put(-250,186){\large 
                     $\frac{\partial G} {\partial \lambda_3}$ 
                   }
    \put(-85,59){ {$\lambda_3$}}
  \end{center}
   {
  \caption{\label{fig:s3} 
     The same as in Fig.~\ref{fig:s1}, but for stage 3.  
     The inset shows the region of the
     plot from $\lambda_3=0.98$ to $0.999$.  
   }}
\end{figure}

The Gibbs free energy change for the i$^{\mbox{th}}$ stage of the path can 
be obtained by
numerical integration: 
\begin{equation}
   \Delta G_{i} =  \int_{0}^{1} d\lambda_i  
                \left( \frac{\partial G}{\partial \lambda_i} \right)
                  =  \int_{0}^{1} d\lambda_i \left\langle 
                \frac{\partial \phi_i}{\partial \lambda_i} \right \rangle, 
\label{eq:dgi}
\end{equation}
where $\langle \cdots \rangle$ represents the isothermal--isobaric ensemble average at a given
value of $\lambda_i$.  The integrands for the various stages of the path
at $1065$ K and zero pressure
are plotted in 
Figs.~\ref{fig:s1}--\ref{fig:s3}. 
It can be seen from these figures that
the value of the integrand for the forward and the
reverse paths agree well. 
This shows that there is no hysteresis present along the path, as found
earlier~\cite{APTE10}.

Throughout this work, we have used the Bennett Acceptance Ratio (BAR) 
method~\cite{BENNETT76},
to compute the Gibbs free energy between the adjacent states along the entire 
path.  According to the BAR method~\cite{BENNETT76}, the Gibbs free 
energy difference $\Delta G=G_1 -G_0$  between two equilibrium 
states `0' and `1', for a given value of the constant $C$, 
is given by the following equation~\cite{BENNETT76}:
\begin {equation}
 \frac{\Delta G}{k_B T} =\log \frac {\sum_1 f(\beta \phi_0-\beta\phi_1+C)}
                    {\sum_0 f(\beta \phi_1-\beta \phi_0-C)}
                       +C - \log \frac{n_1}{n_0} ,
\label{eq:BAR1}
\end {equation}
where 
$f(x)=1/(1+e^x)$ is the Fermi function, 
$\sum_0$ and $\sum_1$ represent the sums over Fermi functions sampled in `0' 
and `1' ensembles, respectively.  The total potential energies in
the two ensembles are represented by $\phi_0$ and $\phi_1$ in
Eq.~(\ref{eq:BAR1}) and  the total
number of samples of the perturbation energies 
(or equivalently the Fermi functions) 
collected
in MC simulations in the two ensembles are 
$n_0$ and $n_1$.
In principle, the
above equation yields the correct value of $\Delta G$ for any value of the
constant $C$.  In practice, due to the limited computational power, 
we cannot sample the perturbation energies
in the two ensembles over all possible configurations.  
Thus, Bennett showed that the optimum value of $C$, which yields minimum
error in the estimation of $\Delta G$,  is given by~\cite{BENNETT76}
\begin {equation}
\frac{\Delta G}{k_B T} = C - \log \frac{n_1}{n_0}.
\label{eq:BAR2}
\end {equation}
Combining the above two equations, we get~\cite{BENNETT76}
\begin {equation}
 {\sum_1 f(\beta \phi_0-\beta\phi_1+C)}
                   = {\sum_0 f(\beta \phi_1-\beta \phi_0-C)}.
\label{eq:BAR3}
\end {equation}
The value of the $C$ satisfying the above equation is substituted in Eq.~(\ref{eq:BAR2})
to yield the optimal estimate of $\Delta G$.
As mentioned in Ref.~\cite{BENNETT76}, the accuracy of the above method depends on
the 
degree of overlap in the configuration space between the two ensembles.  The larger the
value of the two sums appearing in Eq.~(\ref{eq:BAR3}), the greater is the configuration
space overlap.  As prescribed by Bennett, the sum values should be much greater 
than unity~\cite{BENNETT76}.

In all our computations, we ensured that the sum values are of the order of $10^5$--$10^6$, 
by (i) performing simulations at $\lambda_i$ values that are sufficiently close to each other
(see Figs.~\ref{fig:s1}--\ref{fig:s3}) and
(ii) performing sufficiently long simulation runs at each of the $\lambda_i$ values.  The
perturbation energies 
[$(\phi_1 - \phi_0)$ or $(\phi_0 - \phi_1)$ in Eq.~(\ref{eq:BAR3})]
required in the BAR method were collected after every MC step.
At $\lambda_1 = 0.0$ in stage 1, we used the properly equilibrated trajectories
of the HDL phases,
as described in the Sec. II, to collect the perturbation energy data.
In stage 1, we performed upto $20$ million MC steps
from $\lambda_1 = 0$ to $0.4$.  
In the region from
$\lambda_1 = 0.4$ (stage 1) to
$\lambda_3 = 0.99$ (stage 3), we performed $0.4$--$0.8$ million MC steps.  
The main source of statistical error was found to be towards the end of stage 3
due to the center of mass motion of the crystal phase as explained
in detail in Ref.~\cite{APTE10}.  To address this problem, we performed
up to $10$ million MC steps from $\lambda_3 =0.992$ to $\lambda_3 = 0.999$
to collect the perturbation energy data.  
In the last interval of stage 3 from $\lambda_3 = 0.999$ to $1$,
there is a large  change in the integrand (due to center of mass motion of 
the crystalline phase)  as seen 
in the inset of Fig.~\ref{fig:s3}.  In order to minimize the statistical error, 
upto $30$ million MC steps were performed at $\lambda_3 = 1$.
We ensured that the value of the sums $\sum_0$ and $\sum_1$ [see Eq.~(\ref{eq:BAR3})]
over the
Fermi function is above $10^5$ for all the intervals towards the end of stage 3.
Further, in order to improve the 
accuracy three to four independent simulation runs were performed 
along the entire path at
all the temperatures.  
The statistical error for the $i$th stage was computed by using
the formula $\sigma_i = \sigma(\Delta G_i)/n_i^{1/2}$, where $\sigma(\Delta G_i)$
is the standard deviation in the $\Delta G_i$ value and $n_i$ is the number
of statistically independent measurements.  The total statistical error was
computed by adding the errors for the individual stages.  

Table~\ref{tab:table1} lists the values of the excess Gibbs
free energy $G^e$ computed at 
various temperatures in the
supercooled region at and above 1060 K at zero pressure.   
The $G^e$ values are
comparable to those calculated by Broughton and Li~\cite{BROUGHTON87} 
($\sim 25.3 \epsilon$ for N = 512 at 1060 K obtained by
linear interpolation of the chemical potential 
data in Table~III of Ref.~\cite{BROUGHTON87}).   Using the
Gibbs Helmholtz equation for the HDL and the crystalline
phases, we get the following relation between the excess quantities
at different temperatures:
\begin{equation}
       \frac{G_{i+1}^e}{T_{i+1}} 
       -\frac{G_{i}^e}{T_{i}} 
   = - \int_{i}^{i+1} dT~~\left( \frac{H^e}{T^2} \right).
\label{eq:new5}
\end{equation}
Here the quantities with a superscript `e' denote excess quantities.  This
equation can be derived by combining the Legendre transformation 
(LT) relation~\cite{CALLEN85}
$G = H - T S$ and the following expression for the entropy,
\begin{equation}
       S = - \left( \frac{\partial G}
          {\partial T} \right)_{P},
\label{eq:new2}
\end{equation}
for the HDL and the crystalline phases.  The LT relation and 
Eq.~(\ref{eq:new2})
are certainly valid for the crystalline phase, since it 
is the stable equilibrium phase.  
Therefore, the validity of Eq.~(\ref{eq:new5}) 
implies that both the LT
relation and Eq.~(\ref{eq:new2}) are applicable for the HDL phase
as well.
In such a case, it is reasonable to expect that the HDL
phase is an equilibrium phase, since its average enthalpy and average
entropy are determined uniquely by specifying T, P, and N 
[according to the LT relation
and Eq.~(\ref{eq:new2})].

In order to check the validity of Eq.~(\ref{eq:new5}), we have listed the
excess enthalpy $H^e$ values in the last column of Table~\ref{tab:table1}.
For a given temperature interval ($T_i, T_{i+1}$), 
the right hand side of Eq.~(\ref{eq:new5})
can be evaluated numerically by following the trapezoidal rule.  
We find that this equation is satisfied for all the temperature intervals
by our data, except for the data involving the two shorter
trajectories at 1060 K:
at $T_{i+1}=1060$ K, the value predicted 
by Eq.~(\ref{eq:new5}) is $G_{i+1}^e=26.01 \pm 0.024$ for the shorter trajectory 
(1060-S1),  when we  
substitute $G_{i}^e = 25.862 \pm 0.024$ obtained by TDI method at $T_i =1065$ K
in Eq.~(\ref{eq:new5}).  As we can
see the value predicted by Eq.~(\ref{eq:new5}) does {\it not} 
agree within error-bars with
that obtained by the TDI method ($26.073 \pm 0.024$)
for the shorter trajectory at 1060 K (1060-S1).  Same is the case for the
other shorter trajectory 1060-S2.
From this analysis, 
it is reasonable to  conclude 
that the HDL phase properties listed in Table~\ref{tab:table1}
correspond to the equilibrium phases, except for
the shorter trajectory averages (1060-S1 and 1060-S2).
These shorter trajectories correspond
to non-equilibrium states at 1060 K and the LT relation is not applicable to 
such states.


\section{Changes in excess and absolute entropies of the HDL phases}

In order to compute
the excess entropy of the equilibrium HDL phases,
we use the LT relation to obtain $S^e = (H^e - G^e)/T$ (see third column of
Table~\ref{tab:table1}).  The
error in the value of $S^e$, thus computed, is due to 
the error in the estimation of $G^e$.   For the HDL phases generated by
the shorter trajectories at 1060 K, the LT relation is not applicable
as discussed above.  
To compute excess entropy for these
non-equilibrium HDL states, we follow
the thermodynamic analysis by Nishioka~\cite{NISHIOKA87-1}.
In a non-equilibrium state, additional parameters ($X_i;i=1,\cdots,r$) may 
be needed
to specify the state of the system at given values of $T$, $P$, and $N$.   
The first order term for the change in the Gibbs free energy
$G$ as one goes from an equilibrium to a non-equilibrium state 
at constant pressure is given by
\begin{equation}
       \delta G=  \left( \frac{\partial G}
          {\partial T} \right)_{P,X_i} \delta T  
       + \sum_{i=1}^{r} \left( \frac{\partial G}
          {\partial X_i} \right)_{T,P,X_{j\ne i}}  
       \delta X_i
\label{eq:new6}
\end{equation}
Since the initial state is an equilibrium state,
\begin{equation}
       \left( \frac{\partial G}
          {\partial X_i} \right)_{T,P,X_{j \ne i}}  = 0,
\label{eq:new7}
\end{equation}
for $i=1,\cdots,r$.  Applying the analysis by Nishioka~\cite{NISHIOKA87-1}
for the changes between equilibrium states,
Eq.~(\ref{eq:new2}) can be expressed as follows.
\begin{eqnarray} 
       -S &=& \lim_{d T \to 0} \left[
                       \frac{G(T+d T, P, X_i^{eq}+d X_i^{eq})-G(T,P,X_i^{eq})}{d T} \right] \\ \nonumber
         &=& \lim_{d T \to 0} \frac{1}{d T} \left[
                       \left( \frac{\partial G}{\partial T} \right)_{P,X_i} d T + 
             \sum_{i=1}^{r} \left( \frac{\partial G}{\partial X_i}\right)_{T,P,X_{j \ne i}} d X_i^{eq} 
                     + ~\cdots~\right] \\ \nonumber
         &=&  \left( \frac{\partial G}{\partial T} \right)_{P,X_i}, 
\label{eqnarray:old8}
\end{eqnarray}
where we used Eq.~(\ref{eq:new7}) to arrive at the last equality.  The quantities $X_i^{eq}$ 
and $X_i^{eq} + d X_i^{eq}$ 
($i=1,\cdots,r$)
in the above equation correspond to the initial and the final equilibrium states,
respectively.
Substituting Eqs.~(\ref{eq:new7}) and~(14) in Eq.~(\ref{eq:new6}), we have
\begin{equation}
       \delta G=  -S~~\delta T .
\label{eq:old9}
\end{equation}
This equation yields the change in the Gibbs free energy (to the first order)
due to the infinitesimally small variations $\delta T$ and $\delta X_i$
associated with 
a change from the initial equilibrium state to a final non-equilibrium state
at constant pressure.  
For a small but finite variation from the
the initial equilibrium 
state at $T_i = 1065$ K
to a non-equilibrium HDL state at $T_{i+1}=1060$ K at zero pressure, 
$S^e$ can then
be approximated [based on Eq.~(\ref{eq:old9})] as follows:
\begin{equation}
   S^e \approx -~\frac{G^e_{i+1}-G^e_{i}}{T_{i+1} - T_{i}}.
\label{eq:fd}
\end{equation}
Equation~(\ref{eq:fd}) yields the 
average slope (i.e. average value of $S^e$)
in the temperature interval 
($T_i$, $T_{i+1}$).  
This approximation becomes more accurate as $T_{i+1} \rightarrow T_i$.
The values of $S^e$ computed by the above equation (for $N = 512$) 
are reported in the first and second
rows of Table~1 for the trajectories 1060-S1 and 1060-S2, respectively. 
From Table 1,
we find that the average value of $S^e$ in the interval (1060 K,1065 K) is 
higher by about $68 k_B$ (after considering the error bars) for 1060-S1
compared to the corresponding value ($\approx 751 k_B$) in the interval (1065 K, 1070 K).  
Thus, as the 
temperature is decreased by $\Delta T = -5$ K from 1067.5 to 1062.5 K
(the mid-points of above two intervals)
the change in the excess entropy is $\Delta S^e \ge 68~k_B$. 

To estimate the change in entropy of the crystalline phase, we evaluated
its constant volume heat capacity at $1060$ K by using the following
relation~\cite{LEBOWITZ67}
\begin{equation}
  \frac{C_V}{N k_B} = \frac{3}{2}+ \frac{\langle (\delta \phi)^2 \rangle}{N(k_B T)^2} ,
\label{eq:crystalcv}
\end{equation}
where the quantity $\delta \phi = \phi - \langle \phi \rangle$ is the instantaneous fluctuation
in the total potential energy of the system.  The average over the potential energy fluctuations
was computed in the constant volume--constant temperature (NVT)  
MC simulations of the crystal phase and substituting this
value in the above equation yields $C_V = 3.7 N k_B$ at $1060$ K.  We neglect the difference between
the heat capacities 
of the crystal phase at constant pressure and at constant volume, i.e., $C_P \approx C_V = 3.7 N k_B$.
Then the change in absolute entropy of the crystal phase can be estimated by the following formula,
\begin{equation}
  \Delta S \approx {C_P} \log \left( \frac{T+\Delta T}{T} \right),
\label{eq:crystalentropy}
\end{equation}
where we assume $C_P$ to be constant over the temperature range of $\Delta T$.  
Using the above equation, we find that
the change in the entropy 
$\Delta S_{\mbox{\small crystal}}$ of the crystal phase is 
approximately $-9 k_B$ (for $N = 512$) when the
temperature is reduced from 1067.5 K to 1062.5 K ($\Delta T = -5$ K).  
Since 
$\Delta S^e = \Delta S_{\mbox{\small HDL}} - \Delta S_{\mbox{\small crystal}} $,
we find that $\Delta S_{\mbox{\small HDL}} \ge 59~k_B$ 
for the change in temperature
$\Delta T = -5$ K 
from 1067.5 K to 1062.5 K, when we consider
the trajectory 1060-S1.  
It is to be noted that the average energy of 1060-S1 
($\langle E \rangle = -899.93~\epsilon$ for $N=512$)
is slightly higher
as compared to that of the equilibrium HDL phase 
($\langle E \rangle = -900.13~\epsilon$)
at 1065 K
(the average energy can be obtained by adding the 
kinetic energy contribution to the average 
potential energy listed
in the fifth column of Table~\ref{tab:table1}).
This also qualitatively indicates that the average entropy of 
the HDL corresponding to the 1060-S1 trajectory
would be higher as compared to the
equilibrium phase at 1065 K.
For the other trajectory 1060-S2, 
we have  $\Delta S_{\mbox{\small HDL}} \ge 49~k_B$ for the 
temperature change
$\Delta T = -5$ K 
from 1067.5 K to 1062.5 K.
Thus our data indicates that the HDL phase 
shows an increase in the
entropy with decrease in temperature, when 
we consider the non-equilibrium
states (1060-S1 and 1060-S2) at 1060 K.   
For the equilibrium phases, the 
entropy-temperature
relation is monotonic, as expected (see third column of 
Table~\ref{tab:table1}).

\section{Summary and Conclusions}

\begin{figure}
  \begin{center}
    \includegraphics[height=9.8 cm,width=10.2cm]{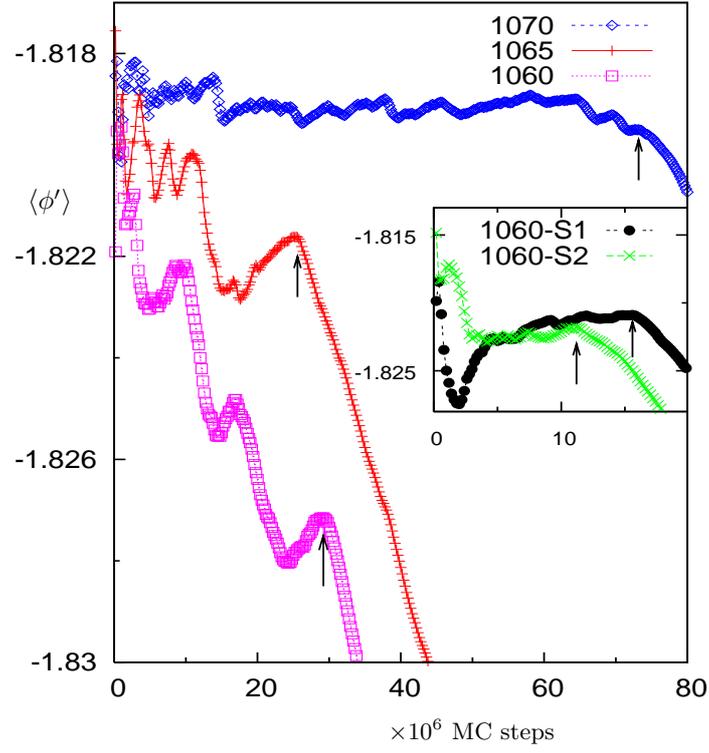}
    \put(-265,190){\normalsize
                     $ \langle \phi' \rangle$
                   }
    \put(-130,-10){$\times 10^6$ MC steps}
  \end{center}
  \caption{\label{fig:trajectory1} 
   The MC trajectories in terms of the cumulative averaged 
   potential energy per particle 
   ($\phi' = \phi/N$ expressed in units of $\epsilon$) at $T=1060$ K, $1065$ K,  and $1070$ K.  
   The points along the trajectories represent cumulative averages taken 
   after every 0.2 million MC steps.  Each MC step, on average,
  consisted of two volume change moves and $N$ particle displacement attempts.
  At 1060 K, the shorter trajectories (1060-S1 and 1060-S2) are 
  shown in the inset, while the longer
  trajectory is shown in the main panel.
    The vertical arrows indicates the
   length of the trajectory used to obtain the average properties of the
   HDL phases.  The trajectories show a systematic and continuous
   decrease in the average energy after the arrow positions, indicating 
   that the free energy barrier is crossed.
   }
\end{figure}

\begin{figure}
  \begin{center}
    \includegraphics[height=9.8 cm,width=10.2cm]{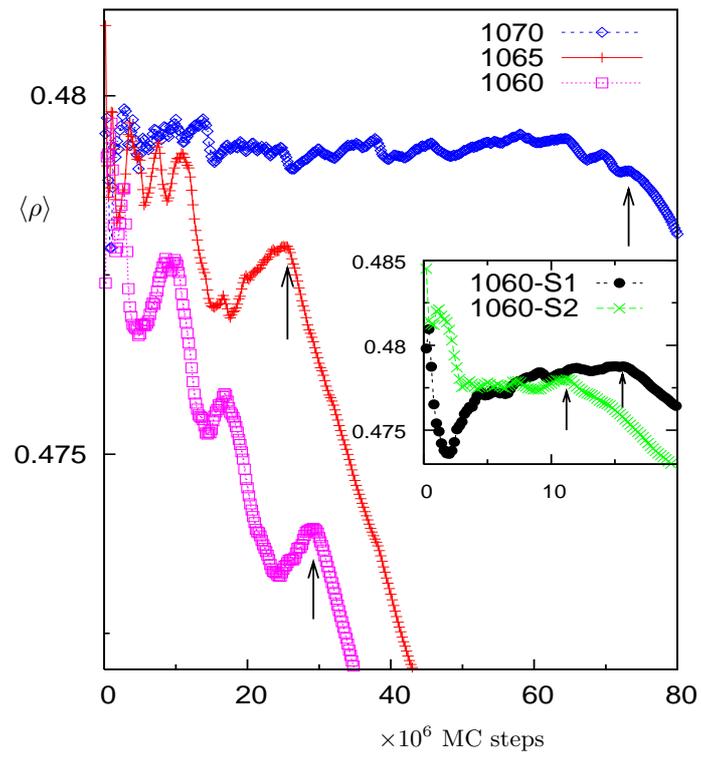}
    \put(-265,190){\normalsize
                     $ \langle \rho \rangle$
                   }
    \put(-130,-10){$\times 10^6$ MC steps}
  \end{center}
  \caption{\label{fig:trajectory2} 
   The same as in Fig.~\ref{fig:trajectory1}, but for the cumulative average
   density (expressed in units of $\sigma^{-3}$).
   }
\end{figure}

\begin{figure}
  \begin{center}
    \includegraphics[height=9.8 cm,width=10.2cm]{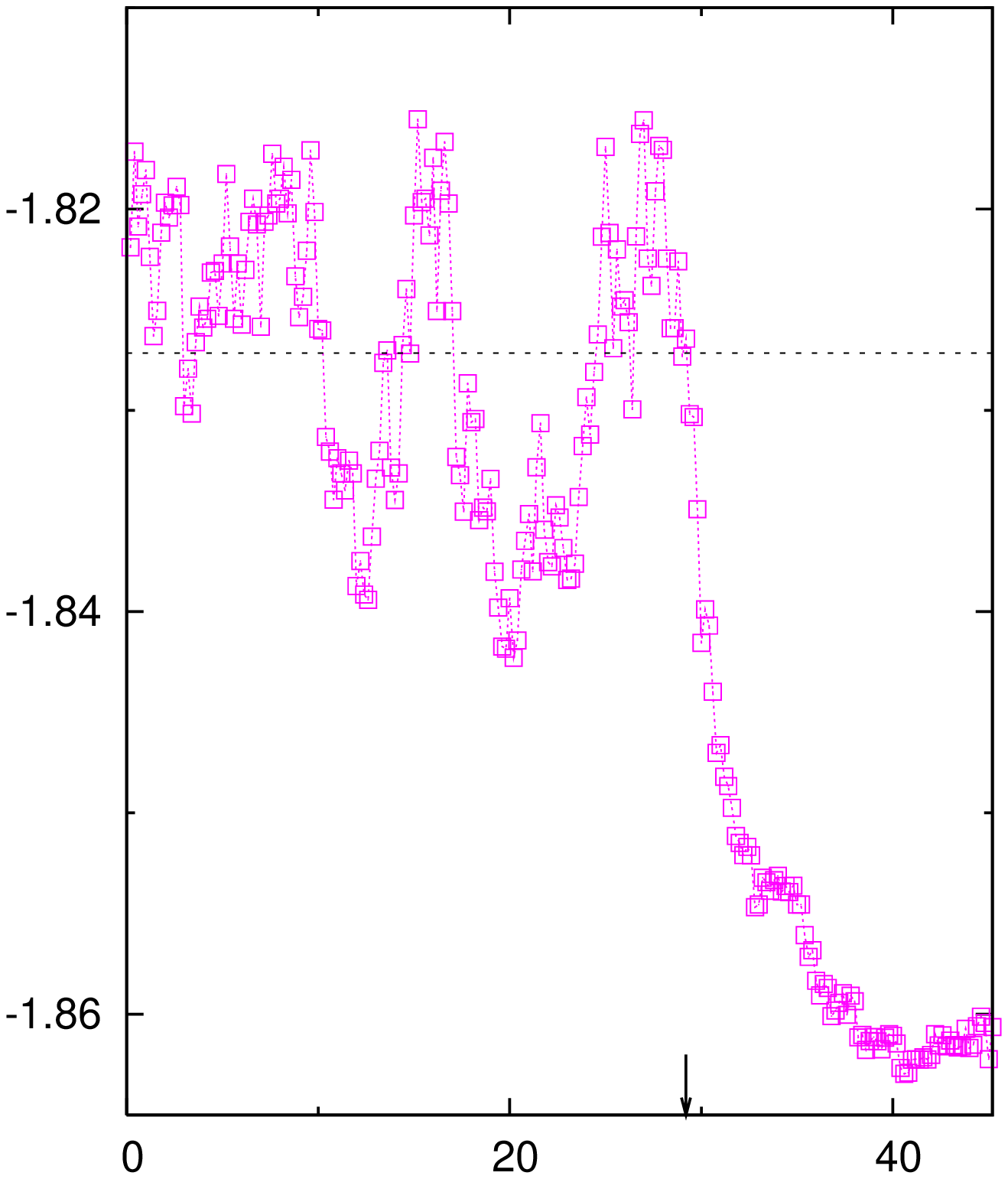}
    \put(-265,190){\normalsize
                     $ \langle \phi' \rangle$
                   }
    \put(-227,85){$\langle \rho \rangle$ }
    \put(-130,-10){$\times 10^6$ MC steps}
  \end{center}
  \caption{\label{fig:trajectory3} 
   The long trajectory at 1060 K (also shown in
   Figs.~\ref{fig:trajectory1}
   and ~\ref{fig:trajectory2}) 
   in terms of {\it instantaneous} 
   block averages taken after every 0.2 million
   MC steps.  The main panel shows average potential energy per particle while the inset
   shows the average density.  The arrow positions (the same as those in earlier figures)
   indicate the location along the trajectory at which 
    the free energy barrier is crossed.
   The horizontal dashed lines shows the cumulative averaged quantities 
  (see earlier figures and also Table~\ref{tab:table1})
  at the arrow positions.  As can be seen in the figure, high energy and high density
  states are accessible  even towards the end of the 
  HDL portion  of the trajectory
  (close to the arrow positions). 
   }
\end{figure}

\begin{figure}
  \begin{center}
    \includegraphics[height=9.8 cm,width=10.2cm]{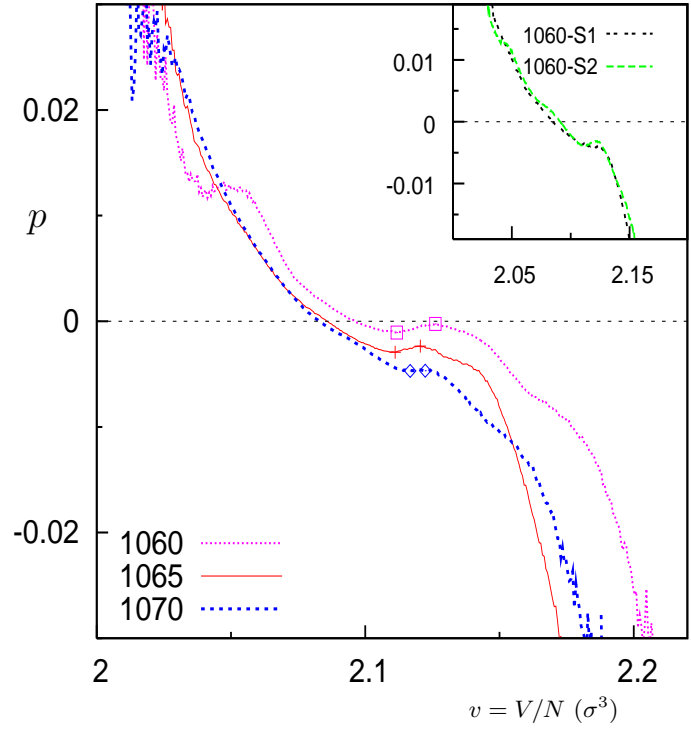}
    \put(-265,180){\Large $p$}
    \put(-100,-5){$v = V/N~(\sigma^{3})$}
  \end{center}
  \caption{\label{fig:spinodal1} 
    Average virial pressure ($p$, in units of $\epsilon/\sigma^3$) 
    as a function of volume per particle ($v=1/\rho$) 
    for the HDL phase.  The
    inset shows the p-v curve for the shorter trajectories at 1060 K.
    At each temperature, the portion of the curve between the two displayed points 
    has a zero or a positive slope indicating a two-phase region. 
    The irregular shape of the curves at the two ends is a result
    of the poor statistical sampling in those regions (see Fig.~\ref{fig:fv1065}). 
   }
\end{figure}

\begin{figure}
  \begin{center}
    \includegraphics[height=9.8 cm,width=10.2cm]{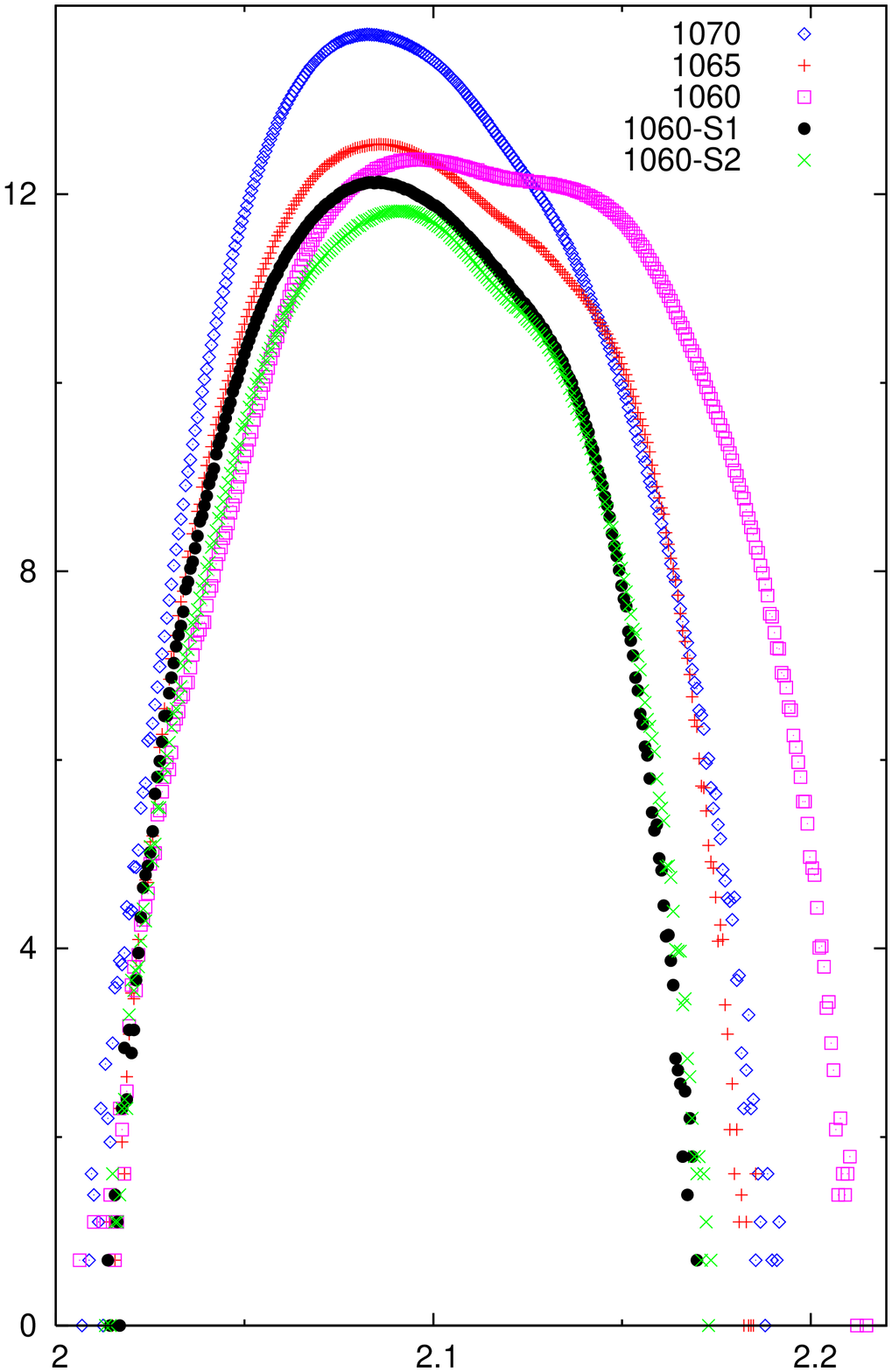}
    \put(-281,180){\begin{sideways} \normalsize{$-\beta F$+const.} 
                   \end{sideways}}
    \put(-100,-10){$v = V/N~(\sigma^{3})$}
  \end{center}
  \caption{\label{fig:fv1065} 
   The Helmholtz free energy (in units of $k_B T$) within a constant 
    as a function of volume for the HDL phases 
   at 1060, 1065, and 1070 K.  As explained in the text, 
   the ordinate is $\log N_c = -\beta F +$ constant, where $N_c$ is
   the number of configurations generated in MC simulations 
   with a specific volume between $v$ and $v+\Delta v$.  
   }
\end{figure}

\begin{figure}
  \begin{center}
    \includegraphics[height=9.6 cm,width=10.2cm]{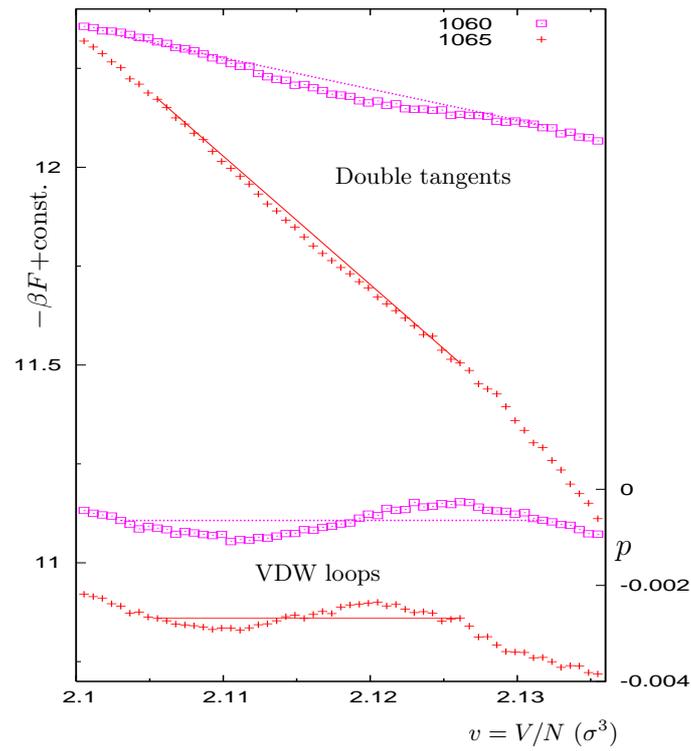}
    \put(-265,145){\begin{sideways} \normalsize{$-\beta F$+const.} 
                   \end{sideways}}
    \put(-45,60){\large $p$}
    \put(-150,200){Double tangents}
    \put(-180,50){VDW loops}
    \put(-100,-10){$v = V/N~(\sigma^3)$}
  \end{center}
  \caption{\label{fig:double_tangent} 
   The double tangent construction at 1060 K and 1065 K.  The upper curves 
   are portions of the
   full F-v curves in Fig.~\ref{fig:fv1065} and represent the
   value of Helmholtz free energy per unit $k_B T$ within a constant.  The lower
   curves are portion of the p-v curves in Fig.~\ref{fig:spinodal1} and 
   represent the average virial pressure as a function of volume per particle.
   The lines in the upper curves are the double tangent lines~\cite{CALLEN85}
   with a slope equal to $\beta p$, where $p$ is the pressure 
   value at the two ends of the
   VDW loops in the lower curves.
   }
\end{figure}

\begin{figure}
  \begin{center}
    \includegraphics[height=9.8 cm,width=10.2cm]{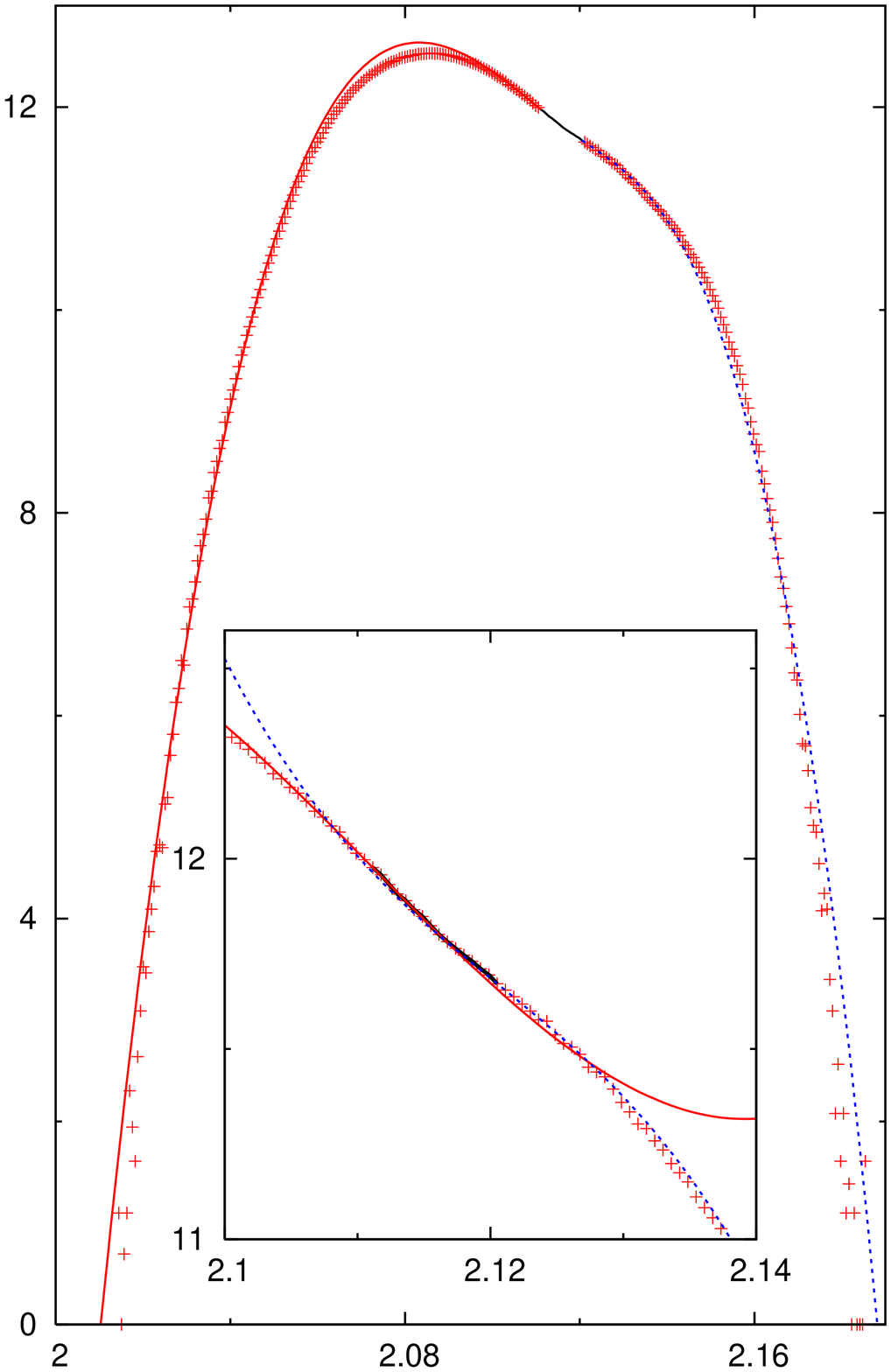}
    \put(-275,183){\begin{sideways} \normalsize{$-\beta F$+const.} 
                   \end{sideways}}
    \put(-100,-10){$v = V/N~(\sigma^{3})$}
  \end{center}
  \caption{\label{fig:sp1065} 
   The red and blue lines denotes the Taylor series expansion to third order around the left and
   right spinodals 
   according to Eq.~(\ref{eq:prob_density3}) at 1065 K.  The lines are truncated
   at the respective spinodals.  The solid black line corresponds to the region of the curve between
   the two spinodals.  The inset shows that the Taylor series
   expansion (red and blue lines) 
   deviate from the actual curve rapidly if continued beyond the spinodals.  
   }
\end{figure}

In this work, we generated NPT MC trajectories corresponding to the
equilibrium HDL phases in the temperature range of $1060$--1070 K
(Figs.~\ref{fig:trajectory1}--\ref{fig:trajectory3}).  
We find
that as the temperature is lowered, the volume distribution becomes
progressively broader and asymmetric (Fig.~\ref{fig:fv1065}), 
exhibiting VDW loops in
the p-v curves (Fig.~\ref{fig:spinodal1} and ~\ref{fig:double_tangent}).
The fluctuations in the HDL phases can access the
low density regions to a greater degree across the VDW loop, upon
lowering the temperature and consequently, the average density
and average energy reduces rapidly.
We were able to construct the double tangent lines (Fig.~\ref{fig:sp1065}) 
across the states forming the
VDW loops at 1060 and 1065 K.  This 
implies that the relation $\Delta U + p \Delta V = T \Delta S$
holds true between these states with $p$ being a negative pressure (Here 
both the terms $\Delta U$ and $p\Delta V$ have the same sign, implying
a non-zero enthalpy difference).
We found that at 1065 K, 
the entire F-v curve (Fig.~\ref{fig:sp1065}) could be described by
a Taylor series expansion around the spinodals 
[Eq.~(\ref{eq:prob_density3})].  

We computed precisely 
the excess Gibbs free energy $G^e$ of the HDL phase
at different temperatures
by using a recently developed thermodynamic integration 
method~\cite{APTE06-1,GROCHOLA04}.  Based on the validity of Eq.~(\ref{eq:new5})
by our data, we found that the HDL phase trajectories correspond to 
the equilibrium phases,
except for the shorter trajectories at 1060 K (1060-S1 and
1060-S2).  We then computed the excess entropy 
of the equilibrium HDL phases from the relation : $S^e = (H^e-G^e)/T$.
For the non-equilibrium HDL phases at 1060 K, we computed $S^e$ from the slope
of $G^e$ with respect to $T$ [Eq.~(\ref{eq:fd})].  
Based on these computations, 
we conclude that the absolute entropy of the HDL phase increases
as its enthalpy changes from the equilibrium value at $T \ge 1065$ K
to the value corresponding to the 
non-equilibrium states at $1060$ K.  
Our conclusion is supported qualitatively
by the fact that the average energy of the HDL phase (1060-S1) 
at 1060 K is slightly higher as compared to that of HDL phase 
at 1065 K (see Table~\ref{tab:table1}).

In the previous MD 
studies~\cite{LUEDTKE89,ANGELL96,BEAUCAGE05} 
it was observed that 
the HDL phase, 
at a sufficiently slow cooling rate, 
transforms into a low density amorphous 
phase at or below $1060$ K.
(Here by amorphous phase, 
we do not necessarily mean the LDL 
phase~\cite{SASTRY03}, 
but any intermediate phase
between the HDL and the LDL phases).  
The trajectories we generated indicate that
the free energy barrier becomes very shallow 
at or below $1060$ K, 
and hence upon cooling
from the high temperature phases ($T \ge 1065$ K), 
it is difficult to achieve the equilibration of the
HDL phases.  
On the other hand, our computations indicate that
the non-equilibrium HDL phases (at $T < 1065$ K) 
one encounters, possess a higher entropy compared to
high temperature equilibrium phases.  
Since the process of {\it cooling} at zero pressure 
necessarily involves reduction of the entropy, 
the HDL phase at $T \ge 1065$ K will
not transform into a higher entropy 
HDL phase at $T < 1065$ K.   
These are the likely reasons which
trigger the transformation of 
the HDL phase into low density
phases near 1060 K upon cooling, as observed
in NPT--MD simulations~\cite{LUEDTKE89,ANGELL96,BEAUCAGE05,HUJO11}.
It is generally supposed that such transformations 
are caused due to 
a coexistence temperature between the HDL and the low density
phases located near 1060 K at zero pressure.  
However, recent studies by Hujo et. al.~\cite{HUJO11} and 
Limmer and Chandler~\cite{LIMMER11} 
do not support this viewpoint.   The volume distributions in the
equilibrium HDL phases at 1060 K and 1065 K
shows low density and high density
states (joined by double tangents 
in Fig.~\ref{fig:double_tangent}) having the same chemical
potential at a {\it non-zero} (negative) pressure.  
Weather such a condition (i.e., 
equality of chemical potentials between
the high density and the low density states) 
is also attainable at zero pressure is
an interesting question and this needs to be investigated further. 

In the case of NPH MD simulations~\cite{SASTRY03}, the
nonmonotonic enthalpy--temperature loop starts at a 
temperature just above $1060$ K.  
Normally, the NPH simulations should trace the equilibrium (monotonic)
enthalpy-temperature curve.  
However, it seems difficult to achieve equilibration of 
the HDL phases
in NPH simulations for the following reasons.
We note that in a statistical mechanical treatment, 
the properties 
of a macroscopic system are not taken as 
strictly constant, 
but are allowed to fluctuate~\cite{dCHANDLER87}.  
This is specially necessary
in the case of HDL phases near 1060 K, which show broad and asymmetric
volume distributions indicating the importance of fluctuations.
In the NPH simulations, on the other hand, the enthalpy is strictly
constant which suppresses relevant fluctuations of enthalpy.  For example,
the two states joined by the double tangents in Fig.~{\ref{fig:double_tangent}} 
have
a non-zero enthalpy difference and both of which contribute significantly
to the average properties.  The NPH simulations cannot access both the states
simultaneously, preventing equilibration of the HDL phases.  This is
the probable reason why the NPH simulations could not access the monotonic
enthalpy--temperature curve corresponding to the equilibrium  HDL phases
at or just above 1060 K~\cite{SASTRY03}.   

The focus of our entire work is an extremely narrow temperature
range 1060-1070 K, though we have also computed HDL phase properties
and excess Gibbs free energies at higher temperature 
(see Table~\ref{tab:table1}).  
We found it difficult to obtain trajectory corresponding to 
an equilibrium HDL phase below 1060 K.  Nonetheless,
our work indicates unexpected but important
changes in the properties of the HDL phase
that are consistent 
with the phase transitions observed in earlier studies at or
near 1060 K.  This, along with the fact that small 
temperature variations
(as low as 0.1 K) are known to induce significant 
changes in the properties 
of other materials~\cite{BUCHANAN12}, justifies our focus on the
narrow temperature range.  

It is desirable to perform free energy computations for the HDL phases
with different number of particles to study system size effects.
However due to
the large extent of computations needed to obtain $G^e$ with a sufficient
precision, it is beyond the scope of the present work.  
It may be noted that the density plots from MD cooling experiment with 
512~\cite{VASISHT11}, 1000~\cite{BEAUCAGE05} and 4096~\cite{HUJO11}
particles are qualitatively similar and show the phase 
transitions near 1060 K.
Recent first principles MD simulations of supercooled  
silicon~\cite{GANESH09}
have demonstrated the 
presence of the VDW loops separating
the high density and low density liquids. 
Thus it is possible that
real silicon exhibits 
phase transition
which is qualitatively similar 
to that of the SW silicon. 

\begin{acknowledgements}
The authors gratefully acknowledge insightful comments by Professor B. D. Kulkarni.
The authors thank Professor Srikanth Sastry and his research group for 
stimulating
discussion and for providing LDL-HDL configurations 
which was helpful for comparison with our MC trajectories.
This work was supported by the young scientist scheme of the 
Department of Science and Technology, India.  
\end{acknowledgements}

\begin{thebibliography}{10}
\providecommand{\url}[1]{{#1}}
\providecommand{\urlprefix}{URL }
\expandafter\ifx\csname urlstyle\endcsname\relax
  \providecommand{\doi}[1]{DOI~\discretionary{}{}{}#1}\else
  \providecommand{\doi}{DOI~\discretionary{}{}{}\begingroup
  \urlstyle{rm}\Url}\fi

\bibitem{ALLEN87}
Allen, M.P., Tildesley, D.J.: {Computer Simulation of Liquids}.
\newblock Oxford University Press, New York (1987)

\bibitem{ANGELL96}
Angell, C.A., Borick, S., Grabow, M.: {Glass transitions and first order
  liquid-metal-to-semiconductor transitions in 4-5-6 covalent systems}.
\newblock J. Non-Cryst. Solids \textbf{205--207}, 463--471 (1996)

\bibitem{APTE10}
Apte, P.A.: {Efficient computation of free energy of crystal phases due to
  external potentials by error--biased Bennett acceptance ratio method}.
\newblock J. Chem. Phys. \textbf{132}, 084,101 (2010)

\bibitem{APTE05}
Apte, P.A., Kusaka, I.: {Direct calculation of solid--liquid coexistence points
  of a binary mixture by thermodynamic integration}.
\newblock J. Chem. Phys. \textbf{123}, 194,503 (2005)

\bibitem{APTE06-1}
Apte, P.A., Kusaka, I.: {Accurate evaluation of translational free energy in a
  melting temperature calculation by simulation}.
\newblock Phys. Rev. E \textbf{73}, 016,704 (2006)

\bibitem{APTE06-2}
Apte, P.A., Kusaka, I.: {Direct calculation of solid--vapor coexistence points
  by thermodynamic integration: Application to single component and binary
  systems}.
\newblock J. Chem. Phys. \textbf{124}, 184,106 (2006)

\bibitem{BEAUCAGE05}
Beaucage, P., Mousseau, N.: {Liquid-liquid phase transition in Stillinger-Weber
  silicon}.
\newblock J. Phys.: Condens. Matter \textbf{17}, 2269--2279 (2005)

\bibitem{BENNETT76}
Bennett, C.H.: {Efficient estimation of free energy differences from Monte
  Carlo data}.
\newblock J. Comput. Phys. \textbf{22}, 245--268 (1976)

\bibitem{BROUGHTON87}
Broughton, J.Q., Li, X.P.: {Phase diagram of silicon by molecular dynamics}.
\newblock Phys. Rev. B \textbf{35}, 9120--9127 (1987)

\bibitem{BUCHANAN12}
Buchanan, M.: {Grain of truth}.
\newblock Nat. Phys. \textbf{8}, 251 (2012)

\bibitem{CALLEN85}
Callen, H.B.: {Thermodynamics and an introduction to thermostatistics}, 2 edn.
\newblock John Wiley \& Sons, New York (1985)

\bibitem{dCHANDLER87}
Chandler, D.: {Introduction to Modern Statistical Mechanics}.
\newblock Oxford University Press, New York (1987)

\bibitem{GANESH09}
Ganesh, P., Widom, M.: {Liquid-liquid transition in supercooled silicon
  determined by first-principles simulation}.
\newblock Phys. Rev. Lett. \textbf{102}, 075,701 (2009)

\bibitem{GROCHOLA04}
Grochola, G.: {Constrained fluid $\lambda$-integration: Constructing a
  reversible thermodynamic path between the solid and liquid state}.
\newblock J. Chem. Phys. \textbf{120}, 2122--2126 (2004)

\bibitem{HUJO11}
Hujo, W., Jabes, B.S., Rana, V.K., Chakravarti, C., Molinero, V.: {The rise and
  fall of anamolies in tetrhedral liquids}.
\newblock J. Stat. Phys. \textbf{145}, 293--312 (2011)

\bibitem{LANDAU-5}
Landau, L.D., Lifshitz, E.M.: {Statistical Physics Part 1}, 3rd edn.
\newblock Pergamon Press, New York (1980)

\bibitem{LEBOWITZ67}
Lebowitz, J.L., Percus, J.K., Verlet, L.: {Ensemble dependence of fluctuations
  with application to machine computers}.
\newblock Phys. Rev. \textbf{153}, 250--254 (1967)

\bibitem{LIMMER11}
Limmer, D.T., Chandler, D.: {The putative liquid-liquid transition is a
  liquid-solid transition in atomistic models of water}.
\newblock J. Chem. Phys. \textbf{135}, 134,503 (2011)

\bibitem{LUEDTKE88}
Luedtke, W.D., Landman, U.: {Preparation and melting of amorphous silicon by
  molecular-dynamics simulations}.
\newblock Phys. Rev. B \textbf{37}, 4656--4663 (1988)

\bibitem{LUEDTKE89}
Luedtke, W.D., Landman, U.: {Preparation, structure, dynamics, and energetics
  of amorphous silicon: A molecular dynamics study}.
\newblock Phys. Rev. B \textbf{40}, 1164--1174 (1989)

\bibitem{NISHIOKA87-1}
Nishioka, K.: {An Analysis of the Gibbs Theory of Infinitesimally Discontinuous
  Variation in Thermodynamics of Interface}.
\newblock Scripta Metallurgica \textbf{21}, 789--792 (1987)

\bibitem{SASTRY03}
Sastry, S., Angell, C.A.: {Liquid-liquid phase transition in supercooled
  silicon}.
\newblock Nat. Mater. \textbf{2}, 739--743 (2003)

\bibitem{STILLINGER85}
Stillinger, F.H., Weber, T.A.: {Computer Simulation of Local Order in Condensed
  Phases of Silicon}.
\newblock Phys. Rev. B \textbf{31}, 5262--5271 (1985)

\bibitem{VASISHT11}
Vasisht, V.V., Saw, S., Sastry, S.: {Liquid--liquid critical point in
  supercooled silicon}.
\newblock Nat. Phys. \textbf{7}, 549--553 (2011)

\end{thebibliography}

%
%
\clearpage
%
%
\begin{center}
\begin{table}
\caption{ \label{tab:table1}
    {The excess Gibbs free energy of the HDL phase with respect to the
     crystalline phase is listed at various temperatures for N = 512 and zero
     pressure.  The excess entropy listed in the third column 
     is computed from the relation $S^e = (H^e-G^e)/T$ for the equilibrium phases.
     The reported error in the $S^e$ value is due to the error in $G^e$ value.
     For the non-equilibrium states (1060-S1 and 1060-S2),
       $S^e$ is calculated by using Eq.~(\ref{eq:fd}).
      The fourth and the fifth columns contain the average
     densities and average potential energy per particle of the HDL phases.
     The excess enthalpy of the HDL phases is listed in the last column of the table.
     The listed values of $H^e$, $S^e$, and $G^e$ correspond to N=512.
    }} 
\begin{tabular}{|c|c|c|c|c|c|} \hline
    {$T$(K)}  &  {$G^e$} ($\epsilon)$       &  $S^e/k_B$ &  
               $\langle \rho  \rangle$ ($\sigma^{-3}$)   &    
               $\langle \phi' \rangle$ ($\epsilon$)      & 
               $H^e$ ($\epsilon$)   \\ 
            &  $N = 512$   &  $N=512$ &  &  &  $N = 512$  \\ \hline
     $1060-\mbox{S1}$ &  $26.073\pm 0.024$   &  $1060 \pm 241 $   & 0.479   & -1.8209 &  $57.753$\\ \hline
     $1060-\mbox{S2}$ &  $26.071\pm 0.024$   &  $1050 \pm 241 $   & 0.478   & -1.8218 &  $57.292$\\ \hline
     $1060$   &  $26.045\pm 0.024   $   &  $ 676.4  \pm 0.6 $         & 0.474   & -1.8272 &  $54.553$\\ \hline
     $1065$   &  $25.862\pm 0.024$   &  $ 740.6 \pm 0.6 $   & 0.478   & -1.8216 & $57.224$ \\ \hline
     $1070$   &  $25.735\pm 0.025$   &  $ 761.6 \pm 0.6 $        & 0.479   & -1.8195 & $58.138$ \\ \hline
     $1075$   &  $25.558 \pm 0.023$   &  $ 771.3 \pm 0.5  $        &  0.479   & -1.8184 & $58.527$ \\ \hline
     $1082$   &  $25.359 \pm 0.029$   &  $ 805.9 \pm 0.7 $        &  0.482   & -1.815  & $60.029$ \\ \hline
     $1086$   &  $25.199 \pm 0.030$   &  $ 815.1 \pm 0.7 $         &  0.482   & -1.814 & $60.398$ \\ \hline
     $1090$   &  $25.088 \pm 0.029$   &  $ 823.4 \pm 0.7 $         &  0.482   & -1.813 & $60.774$ \\ \hline
     $1100$   &  $24.749 \pm 0.028$   &  $ 850.5 \pm 0.6 $           & 0.483   & -1.810  & $61.947$ \\ \hline
\end{tabular}
\end{table}
\end{center}
%
\clearpage
\clearpage
\clearpage
\clearpage
\end{document}